\documentclass[prl,twocolumn,amsmath,amssymb,showpacs,aps]{revtex4}
\usepackage{graphicx}
\usepackage{color}

\begin{document}

\title{Universal Method for Separating Spin Pumping from Spin Rectification Voltage of Ferromagnetic Resonance}
\author{Lihui Bai, P. Hyde, Y.S. Gui, and C.-M. Hu}\email{hu@physics.umanitoba.ca}
\affiliation{Department of Physics and Astronomy, University of Manitoba, Winnipeg, Canada R3T 2N2}

\author{V. Vlaminck\footnote{Current affiliation: Colegio de Ciencias e Ingenieria, Universidad San Francisco de Quito, Quito, Ecuador}, J. E. Pearson, S. D. Bader, and A. Hoffmann}
\affiliation{Materials Science Division, Argonne National Laboratory, Argonne, Illinois 60439, USA}
\affiliation{Nanoscience and Technology Division, Argonne National Laboratory, Argonne, Illinois 60439, USA}
\date{\today}

\begin{abstract}
We develop a method for universally resolving the important issue of separating spin pumping (SP) from spin rectification (SR) signals in bilayer spintronics devices. This method is based on the characteristic distinction of SP and SR, as revealed in their different angular and field symmetries. It applies generally for analyzing charge voltages in bilayers induced by the ferromagnetic resonance (FMR), independent of FMR line shape. Hence, it solves the outstanding problem that device specific microwave properties restrict the universal quantification of the spin Hall angle in bilayer devices via FMR experiments. Furthermore, it paves the way for directly measuring the nonlinear evolution of spin current generated by spin pumping. The spin Hall angle in a Py/Pt bilayer is thereby directly measured as 0.021$\pm$0.015 up to a large precession cone angle of about 20$^{\circ}$.
\end{abstract}

\maketitle

As a promising technique for generating pure spin current in ferromagnetic metal(FM)/normal metal(NM) devices, the intriguing physics of transporting non-equilibrium magnetization pumped by the ferromagnetic resonance (FMR) \cite{Silsbee1979} has received renewed interest. The effect is highlighted under the new concept of spin pumping \cite{Tserkovnyak2005RMP}. In the first three transport experiments on spin pumping performed in 2005 $\sim$ 2006, it was qualitatively understood that spin pumping generates a FMR voltage \cite{Azevedo2005JAP,Saitoh2006APL,Costache2006PRL}. Soon after, a consensus was formed that quantitatively, such FMR voltages in FM/NM bilayers involve in general not only the contribution from spin pumping (SP), but also that from spin rectification (SR), where the magnetization dynamics driven by either the microwave field \cite{Gui2007PRL} or the spin transfer torque \cite{Liu2011PRL} rectifies the microwave current flowing in the FM layer. In 2010, a method based on line shape analysis of the FMR voltage was established for quantitatively separating the two contributions of SP and SR \cite{Mosendz2010PRL}. Although such a line shape analysis is useful and enabled the first quantification of the spin Hall angle via FMR measurement \cite{Mosendz2010PRL}, as discussed in a few follow-up studies \cite{Azevedo2011PRB,Harder2011PRB,Bai2013APL}, it is important to be aware that it only applies on specially designed devices measured under proper configurations which fulfil three conditions: (1) Microwave currents in the non-magnetic metallic layer should be minimized so that contributions from spin transfer torque induced spin rectification can be neglected \cite{Liu2011PRL}. (2) The measurement configuration and microwave phase should be such that the spin rectification either makes no contribution to the Lorentzian part of the electrically detected FMR line shape or can be calibrated \cite{Azevedo2011PRB,Harder2011PRB,Bai2013APL}. (3) The cone angle of the magnetization precession should be small so that the FMR line shape is free from nonlinear distortion \cite{Gui2009}. These strict conditions limit the broad application of the line shape method for generally analyzing the spin pumping and spin Hall effect in bilayer spintronic devices \cite{Hoffman2013IEEE,Liu2011arXiv}. So far, developing a universal method independent of device-specific microwave properties remains a significant challenge. In particular, there is no applicable method for quantifying the spin pumping effect in the nonlinear regime, where the technologically important question of how efficient a large spin current may be generated by high power microwaves remains open.

In this letter, we establish such a universal method based on general symmetry consideration. This method enables the pure spin pumping signal in Py/Pt be directly and unambiguously measured up to the nonlinear regime. For spin pumping up to a large precession cone angle of about 20$^{\circ}$, the spin Hall angle of Pt is measured as a constant of 0.021$\pm$0.015. In contrast, the spin current generated via the nonlinear spin pumping is found to saturate at high pumping powers. Our method brings new insight on spin pumping in the intriguing nonlinear dynamics regime of metallic bilayers.

\begin{figure}[ht]
\includegraphics[width =6 cm]{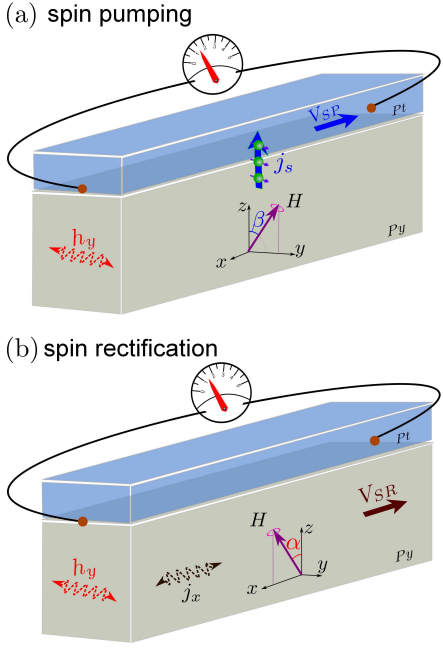}
\caption{ (Color online) Sketch for the DC voltage induced in the FM/NM bilayer by (a) pure spin pumping and (b) pure spin rectification, which can be measured by tilting the direction of the external magnetic field slightly towards the $y$ and $x$ axis, respectively.}
\label{sketch}
\end{figure}

We begin by pointing out the relation of dynamic spintronics responses with the symmetry considerations coined in two classical models \cite{Jan,Silsbee1979}. In particular, spin rectification \cite{Gui2007PRL} dynamically generates a dc voltage $V_{\textmd{SR}}$ via the anisotropic magnetoresistance (AMR), which roots on the broken rotational invariance of FM as revealed in the two-band model \cite{Jan} for spin transport. In contrast, spin pumping \cite{Tserkovnyak2005RMP} produces a dc spin current $\textbf{J}_{s}$ flowing perpendicular to the interface of FM/NM, where the dynamic spin transport property is determined by the breaking of translational invariance at the interface, as highlighted by Silsbee \textit{et al.} in the two-current model \cite{Silsbee1979}. Because of the spin-orbit coupling in the metal \cite{Mosendz2010PRL,Azevedo2011PRB}, $\textbf{J}_{s}$ with the polarization vector of $\vec{\sigma}$ leads to the lateral dc spin pumping voltage $V_{\textmd{SP}}\propto \textbf{J}_{s}\times\vec{\sigma}$. Such a different symmetry relation implies that $V_{\textmd{SR}}$ and $V_{\textmd{SP}}$ can be distinguished by their principle difference in the underlying symmetry breaking mechanisms, which we demonstrate in this letter for a Py/Pt device.

As shown in Fig. \ref{sketch}, let us consider the Py/Pt bilayer carrying a microwave current ($j_{x}$) along the longitudinal $x$ axis, and we choose the $z$ axis as perpendicular to the interface. In such a general device configuration, the dc voltage $V_{x}$ measured longitudinally along the $x$ axis at the FMR frequency $\omega_{r}$ involves both $V_{\textmd{SP}}$ and $V_{\textmd{SR}}$ \cite{Harder2011PRB}. Using $\textbf{e}_{x}$ and $\textbf{e}_{z}$ as the unit vector of the $x$- and $z$-axis, respectively, we find that
\begin{subequations}
\begin{align}
V_{SP} &\propto (|\textbf{m}\cdot\textbf{e}_{H}|\omega_{r})\textbf{e}_{x}\cdot(\textbf{e}_{z}\times\textbf{e}_{H}),    \\
V_{SR} &\propto \langle(\textbf{m}\cdot\textbf{e}_{x})j_{x}\rangle~\textbf{e}_{x}\cdot\textbf{e}_{H}.
\end{align}
\label{eq_symmetry}
\end{subequations}
Here, $\textbf{m}$ = $\textbf{M}(t)$ - $\textbf{M}$ is the non-equilibrium magnetization pumped by the FMR, which makes the saturation magnetization $\textbf{M}$ of the Py deviating from the direction of the externally applied magnetic field $\textbf{H}$ denoted by the unit vector $\textbf{e}_{H}$. Note that in the context of Silsbee \textit{et al.}'s model \cite{Silsbee1979}, $|\textbf{m}\cdot\textbf{e}_{H}|$ in Eq. \ref{eq_symmetry}a accounts for the steady state spin accumulation pumped by FMR, which is the source of the DC spin current $\textbf{J}_{s}$ flowing from Py into Pt due to the broken translational invariance at the interface. In Eq. \ref{eq_symmetry}b, $\langle(\textbf{m}\cdot\textbf{e}_{x})j_{x}\rangle$ is the time average of the product of the oscillating $m_{x}$ and $j_{x}$, which stems from the galvanomagnetism due to the broken rotational invariance in FM. Hence, Eq. \ref{eq_symmetry} highlights the principle difference in the symmetry breaking mechanisms which lead to $V_{\textmd{SP}}$ and $V_{\textmd{SR}}$. It forms the universal ground for separating $V_{\textmd{SP}}$ from $V_{\textmd{SR}}$ independent of device specifics, theoretical parameters, and the line shape of FMR.

Our samples include a Py(15 nm)/Pt(15 nm) bilayer and a reference Py(15 nm) monolayer. Both were deposited on an un-doped GaAs substrate. The structures have a lateral dimension of 20 $\mu$m $\times$ 3 mm and were covered by a signal line (a 30-$\mu$m-width Au) of a coplanar wave guide (CPW) with a MgO layer in between, which provides a microwave magnetic field ($h_{y}$) polarized along the transversal $y$-axis as shown in Fig. \ref{sketch}. By applying a static magnetic field {\bf H} nearly along the $z$ axis, the DC voltage $V_{x}$ is measured at FMR using a lock-in technique by modulating the microwave power at a frequency of 8.33 kHz. Equation \ref{eq_symmetry} indicates that $V_{x}$ has different angular dependence when $\textbf{e}_{H}$ is tilted with an angle of $\alpha$ and $\beta$ towards $\textbf{e}_{x}$ and $\textbf{e}_{y}$, respectively. In particular, it predicts that by setting $\alpha$ and $\beta$ at zero, respectively, pure spin pumping $V_{\textmd{SP}}$ and pure spin rectification $V_{\textmd{SR}}$ signal can be directly detected, which obey the following angular symmetry, respectively:
\begin{align}
\nonumber
At~\alpha =0,~~V_{\textmd{SP}}(\beta, H)& = -V_{\textmd{SP}}(\beta, -H) = -V_{\textmd{SP}}(-\beta, H);\\
At~\beta =0, ~~V_{\textmd{SR}}(\alpha, H)& =~~ V_{\textmd{SR}}(\alpha, -H) = -V_{\textmd{SR}}(-\alpha, H).
\label{eq_VH}
\end{align}
Equation \ref{eq_VH} involves no theoretical parameters, yet it remains unambigius for experimental verification as we demonstrate in our measurements.

\begin{figure}[t]
\includegraphics[width =8.10 cm]{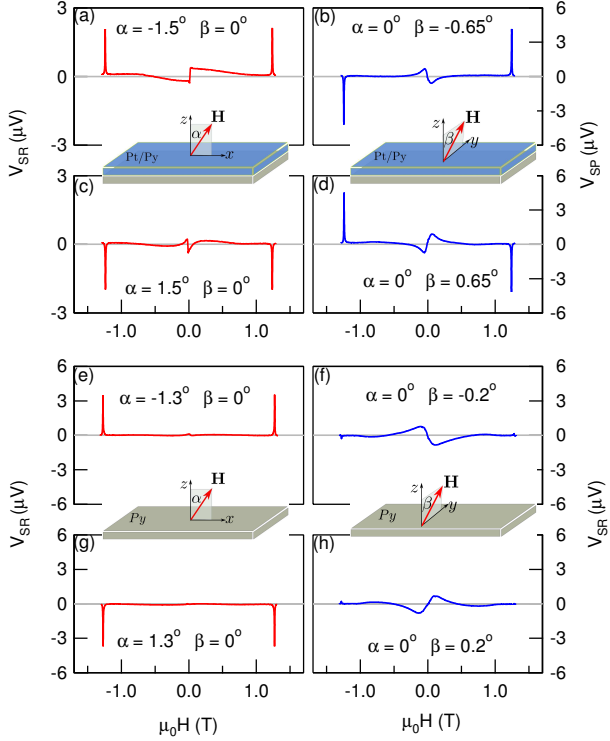}
\caption{ (Color online) DC voltage measured on the Py/Pt bilayer as a function of the magnetic field $H$ applied with angles (a) $\alpha =  -1.5^{\circ} $ and $\beta  = 0$, (b) $\beta  = - 0.65^{\circ}$ and $\alpha = 0$, (c) $\alpha = 1.5^{o} $ and $\beta  = 0$, (d) $\beta  =  0.65^{\circ}$ and $\alpha = 0$, which reveal the distinct symmetries of $V_{SR}$ and $V_{SP}$. (e)-(h) show the comparative results measured on the Py monolayer sample. The microwave frequency is fixed at 9 GHz with a low power of 31.6 mW.}
\label{fig_symmetry}
\end{figure}

Figure \ref{fig_symmetry} highlights the angular ($\alpha$, $\beta$) and field ($H$) symmetry of $V_{x}$ measured in both samples. The microwave sent to the CPW is set at $\omega/2\pi$ = 9 GHz with a low output power of $P$ = 31.6 mW. The FMR appear as the sharp peaks at the fields of $\mu_{0}H_{R}$ = $\pm$1.3 T. As shown in Fig. \ref{fig_symmetry}(a), when $\beta$ is set to zero, the DC voltage of FMR measured at $\alpha = -1.5^{\circ}$ on the Py(15 nm)/Pt(15 nm) bilayer is symmetric about the {\bf H}-field, \textit{i.e.}, $V(H_{R}) = V(-H_{R})$. The polarity of the voltage reverses when $\alpha$ is set to +1.5$^{o}$ as shown in Fig. \ref{fig_symmetry}(c). Hence, the observed symmetry follows exactly what is predicted by Eq. \ref{eq_VH} for the pure spin rectification signal $V_{\textmd{SR}}$. In contrast, as shown in Fig. \ref{fig_symmetry}(b), when $\alpha$ is set to zero, the FMR voltage measured at $\beta = -0.65^{\circ}$ is antisymmetric about the {\bf H}-field, \textit{i.e.}, $V(H_{R}) = -V(-H_{R})$. Again, the polarity reverses when $\beta$ changes its sign, as shown in Fig. \ref{fig_symmetry}(d). Such characteristics follow the prediction of Eq. \ref{eq_VH} for $V_{\textmd{SP}}$. Hence, we identify the voltage measured at $\alpha$ = 0$^{\circ}$ as the pure spin pumping signal. Our conclusion is further confirmed by the results obtained from the reference Py monolayer sample. As expected, pure spin rectification signal $V_{\textmd{SR}}$ measured at $\beta$ = 0$^{\circ}$ preserves, as shown in Fig. \ref{fig_symmetry}(e) and (g), while the pure spin pumping signal $V_{\textmd{SP}}$ measured at $\alpha$ = 0$^{\circ}$ vanishes, as shown in Fig. \ref{fig_symmetry}(f) and (h).

\begin{figure}[t]
\includegraphics[width = 8.10 cm]{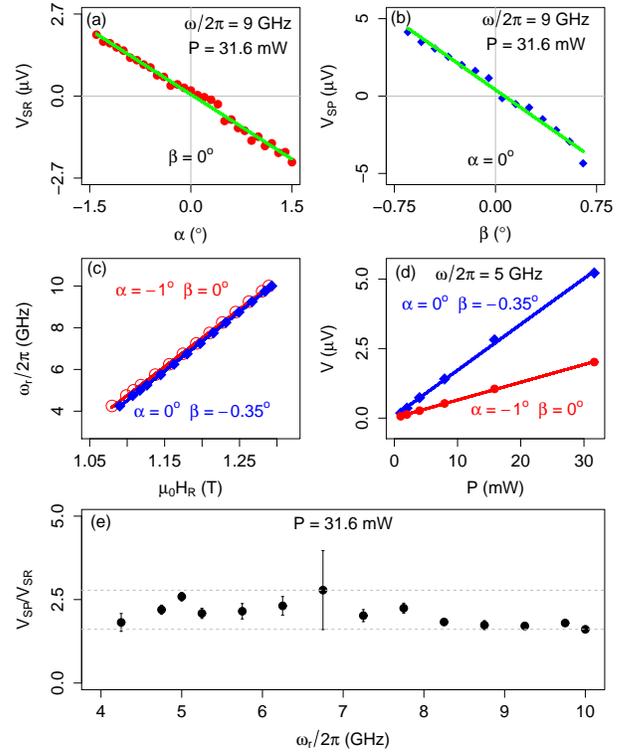}
\caption{ (Color online) Angular dependence of (a) $V_{SR}$ and (b) $V_{SP}$ measured at $\beta$ = 0$^{\circ}$ and $\alpha$ = 0$^{\circ}$, respectively. (c) FMR $\omega_{r}-H_{R}$ dispersions for both pure spin pumping (blue squares) and field torque rectification (red circles). (d) Dependence of $V_{SR}$ and $V_{SP}$ on microwave power measured in the linear dynamic regime up to  $P$ = 31.6 meV.  (e) Frequency dependence of the ratio between $V_{SP}$ and $V_{SR}$ measured at the same microwave power. Here, $V_{SP}$ and $V_{SR}$ are measured at $\alpha = -1^{\circ}$ and $\beta = -0.35^{\circ}$, respectively.}
\label{fig_amp}
\end{figure}

The characteristic results of Fig. \ref{fig_symmetry} are rendered from a large amount of data measured systematically. Figure \ref{fig_amp} plots the data of the bilayer sample, which summarizes the detailed angular, microwave power, and frequency dependence. The dependence of $V_{x}$ on  $\alpha$ and $\beta$ as plotted in Fig. \ref{fig_amp}(a) and (b), respectively, further verifies the prediction of Eq. \ref{eq_VH}. The $\omega_{r}-H_{R}$ dispersion for the FMR is measured at two sets of fixed angles and is plotted in Fig. \ref{fig_amp}(c), both are well fitted by Kittel's formula \cite{Morrish} (solid lines). Note that in the linear dynamic regime, $|\textbf{m}\cdot\textbf{e}_{H}|\propto|m_{x}|^{2}$ and $m_{x}\propto \chi_{xy}\sqrt{P}$, where $\chi_{xy}$ is off-diagonal element of the susceptibility tensor \cite{Nikolai2007PRB}. Hence, Eq. \ref{eq_symmetry} is explicit for the dependence of the DC voltage on both the microwave power $P$ and the FMR frequency $\omega_{r}$. Since at FMR, $|\chi_{xy}|\propto 1/\omega_{r}$, it is straightforward to prove that Eq. \ref{eq_symmetry} indicates that $V_{\textmd{SP}}\propto V_{\textmd{SR}}\propto P/\omega_{r}$ in the linear regime, \textit{i.e.}, both voltages follow the same power and frequency dependence. The measured results plotted in Fig. \ref{fig_amp}(d) confirm that both $V_{\textmd{SP}}$ and $V_{\textmd{SR}}$ are linearly proportional to $P$ (up to 31.6 mW). Keeping $P$ at 31.6 mW, the measured ratio of $V_{\textmd{SP}}/V_{\textmd{SR}}$ as plotted in Fig. \ref{fig_amp}(e) is a constant for different $\omega_{r}$, which confirms that both $V_{\textmd{SP}}$ and $V_{\textmd{SR}}$ follow the same frequency dependence. It also indicates the absence of frequency dependent phase mixing between e and h field in this sample.

\begin{figure}[t]
\includegraphics[width = 8.10 cm]{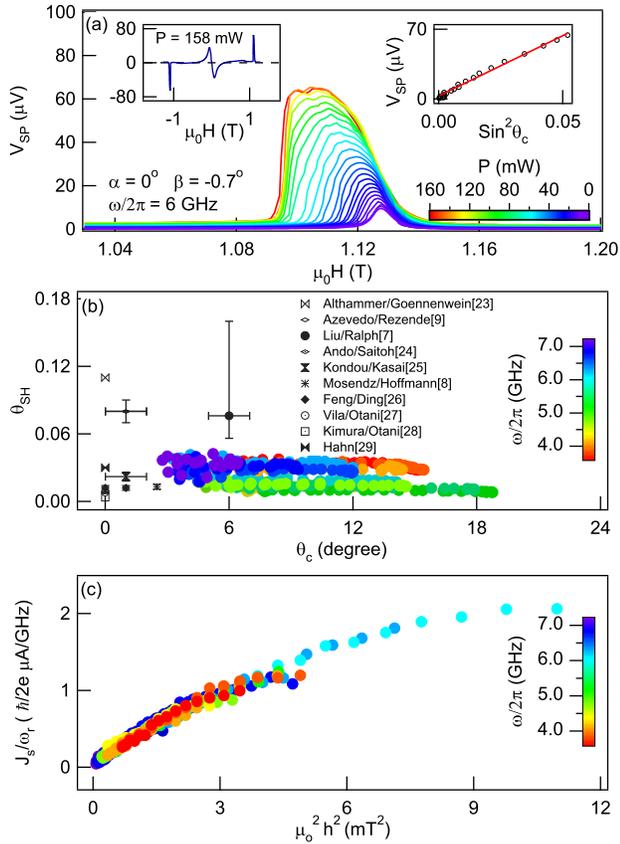}
\caption{ (Color online) (a) $V_{SP}$ measured at $\alpha$ = 0$^{\circ}$ and $\beta$ = -0.7$^{\circ}$ under different microwave excitation power (at a frequency $\omega/2\pi$ = 6.0 GHz). The insets show the full range spectra measured at $P$ = 158 mW, and the dependence of the pure spin pumping voltage on the cone angle. (b) Spin Hall angle of Pt calculated for different cone angles and different frequencies. Results from other experiments are plotted for comparison. (c) Spin current amplitude measured as a function of the microwave magnetic field $h$ and frequency $\omega$. It saturates at high microwave power.}
\label{fig_nonlinear}
\end{figure}

Thus, a general approach based on symmetry analysis is established for separating SP from SR, which is independent of device specifics and theoretical parameters. Indeed, we have tested this method on devices with different structures, with bilayers made of different materials under either in-plane \cite{Mosendz2010PRL,Obstbaum2013}, out-of-plane \cite{Bai2013APL,Rousseau2012PRB,Obstbaum2013} or even arbitrary configuration of microwave excitations \cite{Hyder2013}. In all cases, both pure SP and pure SR signals are obtained. Hence, our method provides for the first time a common ground for different groups to compare their measurements of the FMR voltage, which is a pivotal step towards resolving the controversy of spin Hall effect.

One additional appealing applications of the method is to investigate pure SP in the nonlinear dynamics regime. Technically, useful spintronics devices utilizing SP may require FMR at large cone angles for generating large spin current. However, due to the foldover effect \cite{Gui2009} in the nonlinear regime, the conventional method of studying $V_{\textmd{SP}}$ via line shape analysis \cite{Mosendz2010PRL,Azevedo2011PRB,Harder2011PRB,Bai2013APL} falls short in FM/NM devices. Consequently, pure nonlinear SP has only been detected in a metal/magnetic insulator junction under parametric excitation \cite{Ando2012}. The effect of nonlinear SP in metallic bilayer devices remains unexplored, which we investigate here using our new approach.

Figure \ref{fig_nonlinear}(a) shows $V_{\textmd{SP}}$ measured at $\alpha$ = 0$^{\circ}$ and $\beta$ = -0.7$^{\circ}$ with the output power $P$ of the microwave generator increasing up to 160 mW. Clearly, because of the nonlinear broadening of FMR, the line shape of the FMR voltage measured at high microwave powers can no longer be fitted by using a linear combination of dispersive and Lorentz components as was done in previous studies \cite{Mosendz2010PRL,Azevedo2011PRB,Harder2011PRB,Bai2013APL}. However, as shown in the left inset of Fig. \ref{fig_nonlinear}(a) for the FMR voltage measured at $P$ = 158 mW, it is remarkable that the angular symmetry of the FMR voltage remains the same as that measured at $P$ = 31.6 mW [see Fig. \ref{fig_symmetry}(b)], despite the fact that its amplitude has increased nearly one order of magnitude. It shows that even at the nonlinear dynamic regime, the FMR voltage measured at $\alpha$ = 0$^{\circ}$ remains purely induced by spin pumping, as predicted by Eqs. \ref{eq_symmetry} and \ref{eq_VH}. This paves the way for quantifying both the spin Hall angle of Pt and the spin current amplitude at the Py/Pt interface via nonlinear spin pumping experiment.

To do so, we first determine precisely the cone angle $\theta_{c}$ from the FMR resonant field $H_R$ measured at each microwave power $P$ and frequency $\omega$, by using the simple relation of $H_{R}(\theta_{c}) = H_R(0)-M\theta_{c}^{2}/2$ (see Ref. \cite{Gui2009}). The right inset of Fig. \ref{fig_nonlinear}(a) shows the evolution of $V_{\textmd{SP}}$ as a function of $\sin^2(\theta_{c})$, which enables the precise verification of the SP model \cite{Tserkovnyak2005RMP,Czeschka2011} by accurately showing that $V_{\textmd{SP}}$ is proportional to $\sin^2(\theta_{c})$ in the high power nonlinear regime. Knowing $\theta_{c}$, the spin Hall angle $\theta_{SH}$ in Pt is calculated \cite{Tserkovnyak2005RMP} from the directly measured $V_{\textmd{SP}}$ without performing any line shape fitting. Here, we have adapted a spin diffusion length of 1.3 nm in Pt \cite{Czeschka2011} and the spin mixing conductance of $3 \times 10^{19}$ m$^{-2}$ \cite{Mosendz2010PRL}. Other sample parameters are also given in Ref. \cite{Mosendz2010PRL}. The result of $\theta_{SH}$ determined from such a nonlinear SP experiment is plotted in Fig. \ref{fig_nonlinear}(b). Within the experimental accuracy, we find a constant spin Hall angle of 0.021$\pm$0.015, in the frequency range of 3.6 to 7.2 GHz and up to $\theta_{c}$ of about 20$^{\circ}$. For comparison, we summarized in Fig. \ref{fig_nonlinear}(b) the spin Hall angle of Pt determined from other experiments performed previously either in the linear FMR regime or via dc transport measurements \cite{Liu2011PRL, Mosendz2010PRL, Feng2012PRB, Ando2008PRL, Kimura2007PRL, Althammer2013PRB, Kondou2012APE, Vila2007PRL, Azevedo2011PRB, Hahn2013PRB}. Note that $\theta_{SH}$ is quite diverse for different samples measured by different groups using different methods, which is an outstanding issue of concern \cite{Hoffman2013IEEE,Liu2011arXiv}. It is highly interesting to check whether the diverse results measured on different samples may be verify and analyze by using the universal method defined here.

Finally, our method also enables one to study the efficiency of spin current generation via nonlinear SP. For such a purpose, we use the revised Anderson-Suhl model for nonlinear FMR \cite{Gui2009} to determine the rf magnetic field $h$ that drives the FMR in the Py at each cone angle $\theta_{c}$, and we convert $V_{\textmd{SP}}$ to the spin current amplitude $j_{s}$ at the Py/Pt interface \cite{Tserkovnyak2005RMP}. The $\omega$ and $h$ dependence of $j_{s}$ is summarized and plotted in Fig. \ref{fig_nonlinear}(c). The result verifies that $j_{s}$ is proportional to the microwave frequency \cite{Tserkovnyak2005RMP}. It also shows that with increasing microwave power, the spin current amplitude saturates in the nonlinear regime. Hence, as a pure spin current source, the power efficiency of SP decreases at large microwave power.

In summary, a universal method based on general symmetry consideration is established for clarifying the convoluted mechanisms of the FMR voltage which have plagued the field since 2006. Both the spin Hall angle of Pt and the spin current amplitude at the Py/Pt interface are quantified in the nonlinear spin pumping experiment. Our method defines a common ground for verifying and comparing pure spin pumping signals measured on different spintronics devices, it enables direct quantifying the spin current generated via spin pumping in the nonlinear dynamic regime, and thereby paving the way for designing efficient device structures.

Work in Manitoba (measurements and analysis) was funded by NSERC, CFI, and URGP grants (C.-M.H.). Work at Argonne and use of the Center of Nanoscale Materials (sample fabrication) was supported by the U.S. Department of Energy, Basic Energy Sciences, under Contract No. DE-AC02-06CH11357. The Argonne team thanks H. Schulthei$\ss$ and R. Winkler for stimulating discussions.


\end{document}